\documentclass[12pt,a4paper]{article}
\usepackage{graphicx}
\usepackage{amsmath}
\usepackage{citesort}
\usepackage{amsfonts}
\usepackage{amssymb}

\usepackage{epsfig}

\newcommand\wtktN%
{{\kern-3.8pt}\widetilde{{\kern3.3pt}N{\kern3.3pt}}{\kern-3.4pt}\rangle}
\newcommand\wtktNN%
{{\kern-3.8pt}\widetilde{\widetilde{{\kern3.3pt}N{\kern3.3pt}}}{\kern-3.4pt}\rangle}

\newcommand\wtbrN%
{\langle{\kern-3.4pt}\widetilde{{\kern3.3pt}N{\kern3.3pt}}{\kern-3.4pt}}
\newcommand\wtbrNN%
{\langle{\kern-3.4pt}\widetilde{\widetilde{{\kern3.3pt}N{\kern3.3pt}}}{\kern-3.4pt}}

\newcommand\wtbrNNN%
{\langle{\kern-3.4pt}\widetilde{\widetilde{\widetilde{{\kern3.3pt}N{\kern3.3pt}}}}{\kern-3.4pt}}

\newcommand{\beq}{\begin{equation}}
\newcommand{\enq}{\end{equation}}

\newcommand{\bea}{\begin{eqnarray}}
\newcommand{\ena}{\end{eqnarray}}

\begin{document}
\hfill PTA/09-003
\bigskip

\begin{center}
{\Large{\bf Modular Integrals in Minimal Super Liouville Gravity}}
\end{center}
\vspace{1.0cm}

{\large V.~Belavin}\footnote{Institute of
Theoretical and Experimental Physics, B.Cheremushkinskaya 25, 117259 Moscow, Russia.}

\vspace{0.2cm}

Laboratoire de Physique Th\'eorique et Astroparticules

Universit\'e Montpellier II

Pl.\ E.\ Bataillon, 34095 Montpellier, France

\vspace{1.0cm}

\textbf{Abstract} The four-point integral of the minimal super
Liouville gravity on the sphere is evaluated numerically. The
integration procedure is based on the effective elliptic
parameterization of the moduli space. The analysis is performed for
a few different gravitational four-point amplitudes. The results
agree with the analytic results recently obtained using the Higher
super Liouville equations of motion.

\section{Introduction}

The continuum formulation of the noncritical string theory is
equivalent to 2D quantum gravity coupled to some critical matter,
i.e., the matter described by a conformal field theory
$\mathcal{M}_{c}$. Simple reaction of conformal theories to the
scaling  of the metric leads to the universal form of the effective
action of the generated gravity, which is called the Liouville gravity
(LG)~\cite{Polyakov1}. Because of the peculiarities of two-dimensional
metric geometry, many technical simplifications immediately come into
play. Thus, LG is perhaps the simplest example of quantum gravity,
but it nevertheless shares the same basic questions of
interpretation and can hence be considered useful and worth
studying. The problem of choosing observables correctly and the
problem of calculating the corresponding correlation functions
are of primary importance in any quantum theory. The field of LG has
experienced considerable progress in recent years. Recently
discovered higher equations of motion (HEM)~\cite{HEM} in LFT have
allowed reaching the four-point level in calculating the
correlation functions in LG~\cite{BZ1,BZ11}. The results were tested
against the calculation in the framework of the relatively
independent approach to 2D quantum gravity usually called
the matrix models (see, e.g.,~\cite{GinMoor} and the references
therein). Moreover, a deeper understanding of the correspondence
between these two approaches was achieved
based on these results~\cite{BZ_MM}, although the complete picture
of the relations between these techniques is still missing.

In the context of string theory applications, the construction of 2D
quantum gravity in superspace is one of the most interesting
questions. The first possible generalization is $N{=}1$
supersymmetry. Here, essential progress was also achieved recently.
The study was motivated by a series of works concerned with super
LFT in which the profound understanding of the properties of the
conformal blocks (which are the basic elements of any
CFT~\cite{BPZ}) was achieved. This allowed ``taking off'' the
necessity of treating the correlations involving only special
degenerate~\cite{Kac} excitations. Another important result was the
discovery of the HEM in super LFT~\cite{superhigher}. All these
results have served as a good starting point for a more profound
study of super Liouville gravity (SLG). In~\cite{BBSLG}, the
structure of the physical fields in the Neveu--Schwarz (NS) sector
of SLG was clarified, and the general expression for the $n$-point
correlation numbers on the sphere in terms of integrals over moduli
space was written explicitly. Then the super HEM in the $N{=}1$
super LFT and the analysis of the structure of the super ground ring
(and its logarithmic counterpart) allowed deriving the explicit
analytic expressions for the four-point correlators. Because little
is now known for the supersymmetric matrix model, there are no
independent results analogous to those obtained in~\cite{BBSLG}. In
this situation, more checks of the validity of these results would
seem desirable. This paper is devoted to directly calculating the
four-point correlation numbers in SMLG.

The paper is organized as follows. To make the presentation
self-contained, we collect all necessary information related to the
subject in Sec.~2 and Sec.~3. The remaining part of the paper deals
with evaluating the four-point integrals numerically. In Sec.~4, we
consider two examples of the four-point integral. We reduce the
expressions to a number of integrals over the fundamental region of
the modular group, and the integrands represent the products of the
various correlation functions both in the matter and in the
Liouville sectors. The elliptic transformation is the subject of
Sec.~5. The numerical results are presented in Sec.~6. Some useful
details omitted from the main text are given in Appendix~A.

\section{ Minimal Super Liouville Gravity}

In the framework of the so-called DDK approach~\cite{D,DK,HDK}, SLG is
represented as a tensor product of superconformal matter (SM), super
Liouville, and super ghost systems
\begin{equation}
A_{\text{SLG}}=A_{\text{SM}}+A_{\text{SL}}+A_{\text{SG}}
\end{equation}
with the interaction via the relation for the central
charge parameters
\begin{equation}
c_{\text{SM}}+c_{\text{SL}}+c_{\text{SG}}=0 \label{totc}
\end{equation}
and also due to the construction of the physical fields.

The superconformal algebra is
\begin{equation}
\begin{aligned}
\lbrack L_n,L_m]&=(n-m)L_{n+m}+\frac{\hat c}8(n^3-n)\delta_{n,-m},
\\
\{G_r,G_s\}&=2L_{r+s}+\frac{\hat
c}2\left(r^2-\frac14\right)\delta_{r,-s},
\\
[L_n,G_r]&=\left(\frac12n-r\right)G_{n+r},
\end{aligned}
\label{The algebra}
\end{equation}
where
\begin{equation*}
\begin{alignedat}{2}
&r,s\:\in\mathbb{Z}+\frac12&\quad&\text{for the NS sector},
\\
&r,s\:\in\mathbb{Z}&\quad&\text{for the R sector}.
\end{alignedat}
\end{equation*}

The SLFT central charge is
\begin{equation}
\widehat{c}_{\text{SL}}=1+2 Q^{2},\label{cQ}%
\end{equation}
where the ``background charge'' parameter $Q$ is related to the
SLFT basic quantum parameter~$b$
\begin{equation}
Q=b^{-1}+b
\end{equation}
The fields belong to the highest-weight representations of the
superconformal algebra. The basic fields of interest in this paper
belong to the primary supermultiplet $(V_{a},Y_a,\bar Y_a, W_a)$
with the bottom component $V_a$ having the conformal dimension
\begin{equation}
\Delta^{\text{L}}_{a}=\frac{a(Q-a)}2,\label{Da}%
\end{equation}
where $a$ is a continuous (complex) parameter, and the other
components of the primary supermultiplet being
\begin{equation}
Y_a=G^{\text{L}}_{-1/2}V_a,\,\,\,\,\,\bar Y_a=\bar
G^{\text{L}}_{-1/2}V_a,\,\,\,\,\,W_a=\bar G^{\text{L}}_{-1/2}
G^{\text{L}}_{-1/2}V_a.
\end{equation}
Here and hereafter, we use the superscripts M, L, and G to specify the
matter, Liouville, and ghost sectors of the superconformal generators.
The generators without sector superscripts are related to the total
super Virasoro algebra.

At certain special values of the parameter $a=a_{m,n}$, one singular
vector appears at the level $mn/2$ in the Verma module over
$V_{a_{m,n}}=V_{m,n}$~\cite{Kac}. Here,
\begin{equation}
a_{m,n}=Q/2-\lambda_{m,n},
\end{equation}
where $(m,n)$ is a pair of positive integers ($m-n\in2\mathbb{Z}$) and
\begin{equation}
\lambda_{m,n}=\frac{mb^{-1}+nb}2.
\label{lmn}%
\end{equation}

The basic super Liouville operator product expansion~\cite{BBNZ}
(for the sake of brevity we write $\Delta=\Delta_{Q/2+iP}$ and
$\Delta_{i}=\Delta_{a_{i}}$)
\begin{align}
\  & V_{a_{1}}(x)V_{a_{2}}(0)=\label{VV}\\
& \ \ \ \ \int' \frac{dP}{4\pi}\left(  x\bar x\right)
^{\Delta-\Delta _{1}-\Delta_{2}}\left(
\mathbb{C}_{a_{1},a_{2}}^{Q/2+iP}\left[
V_{Q/2+iP}(0)\right]  _{\text{ee}}+\mathbb{\tilde C}_{a_{1},a_{2}}%
^{Q/2+iP}\left[  V_{Q/2+iP}(0)\right]  _{\text{oo}}\right) \nonumber
\end{align}
This OPE is continuous and involves integration over the
``momentum'' $P$. In (\ref{VV}), $\left[V_{p}\right]_{\text{ee,oo}}$
denotes the contribution of the primary field $V_{p}$ and its
``even'' and ``odd'' superconformal descendants to the operator
product expansion. As usual, the prime on the integral indicates
possible discrete terms; in this study, we consider only the region
$b$ where such extra terms do not appear and the integral can be
understood literally. All other OPEs of two arbitrary local fields
in SLFT can be derived from~(\ref{VV}). The basic structure
constants $\mathbb{C}_{a_{1}a_{2}}^{Q/2+iP}$ and
$\tilde{\mathbb{C}}_{a_{1},a_{2}}^{Q/2+iP}$ in~(\ref{VV}) were
evaluated using the bootstrap technique in~\cite{Rubik,Marian} and
have the explicit form (here $a$ denotes $a_{1}+a_{2}+a_{3}$)
\begin{equation}
\begin{aligned}
\mathbb{C}_{a_{1}a_{2}}^{Q-a_3}&=\left(\!\pi\mu\gamma\!
\left(\frac{Qb}2\right)b^{1-b^2}\right)^{\!\!(Q-a)/b}\!
\frac{\Upsilon_{\text{R}}(b)\Upsilon_{\text{NS}}(2a_{1})
\Upsilon_{\text{NS}}(2a_{2})\Upsilon_{\text{NS}}(2a_{3})}
{2\Upsilon_{\text{NS}}(a-Q)\Upsilon_{\text{NS}}%
(a_{1+2-3})\Upsilon_{\text{NS}}(a_{2+3-1})\Upsilon_{\text{NS}}(a_{3+1-2})},
\\
\tilde{\mathbb{C}}_{a_{1}a_{2}}^{Q-a_3}&=-\left(\!\pi\mu\gamma\!
\left(\frac{Qb}2\right)b^{1-b^{2}}\right)^{\!\!(Q-a)/b}
\frac{i\Upsilon_{\text{R}}(b)\Upsilon_{\text{NS}}(2a_{1})
\Upsilon_{\text{NS}}(2a_{2})\Upsilon_{\text{NS}}(2a_{3})}
{\Upsilon_{\text{R}}(a-Q)\Upsilon_{\text{R}}%
(a_{1+2-3})\Upsilon_{\text{R}}(a_{2+3-1})\Upsilon_{\text{R}}(a_{3+1-2})},
\end{aligned}
\label{C3}
\end{equation}
where we use the convenient notation in~\cite{Fukuda} for the
special functions
\begin{equation}
\begin{aligned}
\Upsilon_{\text{NS}}(x)&=\Upsilon_{b} \left(\frac
x2\right)\Upsilon_{b}\left(\frac{x+Q}2\right),
\\
\Upsilon_{\text{R}}(x)&=\Upsilon_{b}
\left(\frac{x+b}2\right)\Upsilon_{b}\left(\frac{x+b^{-1}}2\right)
\end{aligned}
\label{YNSR}
\end{equation}
expressed in terms of the ``upsilon'' function $\Upsilon_{b}$, which
is the standard element in the Liouville field theory
(see~\cite{DO,LFT}).

Because of central charge balance condition~\eqref{totc}, the central
charge of the matter sector is given in terms of the same basic
parameter $b$:
\begin{equation}
\widehat{c}_{\text{SM}}=1-2q^{2},\label{cQ}%
\end{equation}
where $q=b^{-1}-b$. We let
$(\Phi_{\alpha},\chi_{\alpha},\bar\chi_{\alpha},\Psi_{\alpha})$
denote the primary multiplet in the matter sector with
the dimension of the bottom component $\Phi_a$ being given by
\begin{equation}
\Delta^{\text{M}}_{\alpha}=\frac{\alpha(q-\alpha)}2.\label{DMa}%
\end{equation}

The super ghost system (see, e.g.,~\cite{Polchinski1,Verlinde,Fridan})
is described by the free super conformal field theory with the central
charge $c_{\text{SG}}=-10$. The fermionic part of the SG system involves
two anticommuting fields $(b,c)$ of spins $(2,-1)$, and the bosonic
part involves two bosonic fields $(\beta,\gamma)$ of spins
$(3/2,-1/2)$. The formal fields (see~\cite{BBSLG}) of the form
$\delta(\gamma(0))$ of dimension $1/2$ are essential in constructing
the gravitational amplitudes.

\section{Physical Fields and the Correlation Numbers}

The physical fields form a space of cohomology classes with respect
to the nilpotent BRST charge $\mathbb{Q}$,
\begin{align}
\mathbb{Q}=\sum_m{:}\bigg[L_m^{\text{M+L}}+\frac{1}{2}L^{\text{g}}_m\bigg]c_{-m}{:}+
\sum_r{:}\bigg[G_r^{\text{M+L}}+\frac{1}{2}G^{\text{g}}_r\bigg]\gamma_{-r}{:}-\frac{1}{4}c_0.
\label{Q}
\end{align}
In this work, we deal with the correlators of physical fields of the two
types
\begin{align}
\mathbb{W}_{a}(z,\bar z)=U_{a}(z,\bar z)\cdot c(z)\bar c(\bar
z)\cdot \delta(\gamma(z))\delta(\bar\gamma(\bar z)), \label{W}
\end{align}
and
\begin{align}
\tilde{\mathbb{W}}_{a}(z,\bar z)=\biggl(\bar G^{\text{M+L}}_{-1/2}+
\frac12\bar G_{-1/2}^{\text{g}}\biggr)\biggl(G^{\text{M+L}}_{-1/2}+
\frac12G_{-1/2}^{\text{g}}\biggr)\mathbb{U}_{a}(z,\bar z)\cdot\bar
c(\bar z)c(z), \label{tildeW}
\end{align}
where
\begin{align}
\mathbb{U}_{a}(z,\bar z)=\Phi_{a-b}(z,\bar z)V_{a}(z,\bar z)
\end{align}
Here the parameter $a$ can take generic values. The general form of
the $n$-point correlation numbers on the sphere for these
observables~\cite{BBSLG} is
\begin{equation}
I_n(a_1,\cdots,a_n)=\prod_{i=4}^{n}\int d^2 z_i \biggl\langle \bar
G_{-1/2}G_{-1/2}\mathbb{U}_{a_i}(z_i)
\tilde{\mathbb{W}}_{a_{3}}(z_{3})\,\mathbb{W}_{a_{2}}(z_{2})\,
\mathbb{W}_{a_{1}}(z_{1})\biggr\rangle. \label{corrSMG}
\end{equation}

An additional ``discrete'' physical state arises when the
representation in the matter sector is degenerate,
\begin{equation}
\mathbb{O}_{m,n}(z,\bar z)=\bar H_{m,n}H_{m,n}\Phi_{m,n}(z,\bar
z)V_{m,n}(z,\bar z).
\end{equation}
The operators $H_{m,n}$ are composed of the super Virasoro
generators and are defined uniquely modulo exact terms. Moreover, if
we introduce the logarithmic counterparts of the discrete states
$\mathbb O_{m,n}$,
\begin{equation}
\mathbb{O}'_{m,n}=\bar H_{m,n}H_{m,n}\Phi_{m,n}V_{m,n}',
\end{equation}
then we have the important relations~\cite{BBSLG}
\begin{equation}
  \bar{\mathbb{Q}} \mathbb{Q} \mathbb{O}'_{m,n}=B_{m,n}\tilde{\mathbb{W}}_{m,-n} \label{basic}
\end{equation}
and
\begin{align}
\bar G_{-1/2}G_{-1/2}\mathbb{U}_{m,-n}=
B_{m,n}^{-1}\bar\partial\partial \mathbb{O}'_{m,n}\mod \mathbb{Q},
\label{GGU1}
\end{align}
where $B_{m,n}$ are the coefficients arising in the higher equations
of motion of SLFT~\cite{superhigher}. For four points,
relation~\eqref{GGU1} allows reducing the moduli integral in
general expression~\eqref{corrSMG} to boundary integrals if one
of the fields is degenerate, i.e., $a_i=a_{m,-n}$. The explicit
result is (see~\cite{BBSLG})
\begin{equation}
I_4(a_{m,-n},a_1,a_2,a_3)=\kappa
N(a_{m,-n})\biggl\{\sum_{i=1}^3\sum_{r,s\in (m,n)}
q_{r,s}^{(m,n)}(a_i)+2 m n\lambda_{m,n}\biggr\}\prod_{i=1}^3 N(a_i),
\label{4point}
\end{equation}
where
\begin{equation}
q_{r,s}^{(m,n)}(a)=|a-\lambda_{r,s}-Q/2|_{\text{Re}}-\lambda_{m,n}
\end{equation}
and the fusion set is $(m,n)=\{1-m:2:m-1,1-n:2: n-1\}$. The coefficient is
\begin{equation}
\kappa=-2 \mu^{-1} b^{-2} \biggl[\pi \mu
\gamma\biggl(\frac12+\frac{b^2}2\biggr)\biggr]^{2+b^{-2}}\gamma\biggl[\frac{b^{-2}}{2}-\frac12\biggr],
\end{equation}
and the ``leg'' factors  are
\begin{equation}
N(a)=\biggl[\pi\mu\gamma\biggl(\frac12+\frac{b^2}2\biggr)\biggr]^{-a/b}
\biggl[\gamma\biggl(ab-\frac{b^2}2+\frac12\biggr) \gamma\biggl(\frac
ab-\frac{b^{-2}}2+\frac12\biggr)\biggr]^{1/2}. \label{N}
\end{equation}

\section{Direct Calculation}

Here, we verify analytic result~\eqref{4point}. The space of
parameters ($a_1$, $a_2$, $a_3$, and also $b$) is rather big to
present a comprehensive analysis. In what follows, we focus on the
two examples where one of the fields is either $\mathbb{W}_{b}$ or
$\mathbb{W}_{2b}$. Moreover, we restrict ourself to considering the
most symmetric situation of four identical fields
$\mathcal{I}_4(a)=I_4(a,a,a,a)$. In the four-point
case~\eqref{corrSMG} reduces to
\begin{align}
\mathcal{I}_{4}(a)= \int d^2 z \langle \bar G_{-1/2} G_{-1/2}
\mathbb{U}_{a}(z) \mathbb{W}_{a}(0)\, \tilde{\mathbb{W}}_{a}
(1)\,\mathbb{W}_{a} (\infty) \rangle. \label{I40}
\end{align}
The analytic results following from expression~\eqref{4point} are
\begin{align}
\mathcal{I}_4(b)=\frac{\kappa}b N^4(b)\Sigma^{(1,1)}(b),
\label{I1analit}
\end{align}
where
\begin{align}
\Sigma^{(1,1)}(b)=|2b^2-1|.
\end{align}
For the second integral, we have
\begin{align}
\mathcal{I}_{4}(2b)=\frac{\kappa}b N^4(2b) \Sigma^{(1,3)}(b)
\label{I2analit}
\end{align}
and
\begin{align}
\Sigma^{(1,3)}(b)=\frac32\bigg[|5b^2-1|+|3b^2-1|+|b^2-1|-3b^2-1\bigg].
\end{align}

Let us consider the integral $\mathcal{I}_4(b)$. Taking into account
that we deal with the unit operators in the matter sector in this
case, we have
\begin{align}
&\bar G_{-1/2} G_{-1/2} \mathbb{U}_{b}=W_b,\nonumber \\
&\mathbb{W}_{b}=V_b \,\bar c c \,\delta(\bar\gamma)\delta(\gamma),\\
&\tilde{\mathbb{W}}_{b} =\bigg(\bar
G_{-1/2}^{\text{M}+\text{L}}+\frac12 \bar
G_{-1/2}^{\text{g}}\bigg)\bigg( G_{-1/2}^{\text{M}+\text{L}}+\frac12
G_{-1/2}^{\text{g}}\bigg) V_b \,\bar c c.\nonumber
\end{align}
Taking the explicit form of the correlation functions in the ghost sector
into account,
\begin{align}\label{ghostcorr}
\langle C(0) C(1) \rangle=0,\nonumber\\
\langle C(0) C(1) C(\infty) \rangle=1,\\
\langle\delta(\gamma(0))\delta(\gamma(1))\rangle=1,\nonumber
\end{align}
we conclude that the only nonzero contribution comes from the term
in $\tilde {\mathbb{W}}_b$, which is proportional to $\bar c c$,
\begin{align}
\mathcal{I}_4(b)=\int d^2 z \langle
W_b(z)V_b(0)W_b(1)V_b(\infty)\rangle.
\end{align}
In the same way, we have
\begin{align}
&\bar G_{-1/2} G_{-1/2} \mathbb{U}_{2b}=\bar
G_{-1/2}^{\text{M}+\text{L}}G_{-1/2}^{\text{M}+\text{L}}\Phi_b V_{2b},\nonumber \\
&\mathbb{W}_{2b}=\Phi_b V_{2b} \,\bar c c \,\delta(\bar\gamma)\delta(\gamma),\\
&\tilde{\mathbb{W}}_{2b} =\bigg(\bar
G_{-1/2}^{\text{M}+\text{L}}+\frac12 \bar
G_{-1/2}^{\text{g}}\bigg)\bigg( G_{-1/2}^{\text{M}+\text{L}}+\frac12
G_{-1/2}^{\text{g}}\bigg)\Phi_b V_{2b} \,\bar c c,\nonumber
\end{align}
for the second integral, and taking~\eqref{ghostcorr} into account,
we obtain
\begin{align}
\mathcal{I}_4(2b)=\int d^2 z \bigg(
\langle\Psi_b(z)\Phi_b(0)\Psi_b(1)\Phi_b(\infty)\rangle
\langle V_{2b}(z)V_{2b}(0)V_{2b}(1)V_{2b}(\infty)\rangle,\nonumber\\
+\langle\chi_b(z)\Phi_b(0)\chi_b(1)\Phi_b(\infty)\rangle \langle\bar
Y_{2b}(z)V_{2b}(0)\bar Y_{2b}(1)V_{2b}(\infty) \rangle,\\
+\langle\bar\chi_b(z)\Phi_b(0)\bar\chi_b(1)\Phi_b(\infty)\rangle
\langle  Y_{2b}(z)V_{2b}(0) Y_{2b}(1)V_{2b}(\infty) \rangle,\nonumber\\
+ \langle \Phi_b(z)\Phi_b(0)\Phi_b(1)\Phi_b(\infty)\rangle \langle
W_{2b}(z)V_{2b}(0)W_{2b}(1)V_{2b}(\infty)\rangle.\nonumber \bigg)
\end{align}

We now use the symmetry of the integrals under modular transformations to
reduce the integration from the whole complex plane to the fundamental
domain. The modular subgroup of projective transformations divides the
complex plane into six regions. The fundamental region is defined as
$\mathbf{G=}\{\operatorname*{Re}x<1/2;\;\left| 1-x\right| <1\}$. The other
five regions are mapped to the fundamental one using one of the
transformations $\mathcal{A},\mathcal{B},\mathcal{A}\mathcal{B},
\mathcal{B}\mathcal{A},\mathcal{A}\mathcal{B}\mathcal{A}$,
where $\mathcal{A}$: $z\rightarrow 1/z$ and $\mathcal{B}$: $z\rightarrow
1-z$. Combining the projective transformations of the fields and the
corresponding change of variables in the integrals, we reduce the
integration to the fundamental region. We note that the Jacobian of the
transformation exactly cancels the transformation of the fields
because their total conformal dimension is $1$. Then,
\begin{align}
\mathcal{I}_4(b)=2\int_{\mathbf{G}} d^2 z \bigg(\langle W_b(z)V_b(0)W_b(1)V_b(\infty)\rangle+
\langle W_b(z)V_b(0)V_b(1)W_b(\infty)\rangle+\nonumber\\
+\langle W_b(z)W_b(0)V_b(1)V_b(\infty)\rangle\bigg),
\end{align}
where the factor $2$ in front of the integral takes the equivalent
projective images into account. The expression for the second integral
is rather bulky:
\begin{align}
\mathcal{I}_4(2b)=2\int_{\mathbf{G}} d^2 z \bigg[\bigg(
\langle\Psi_b(z)\Phi_b(0)\Psi_b(1)\Phi_b(\infty)\rangle \langle
V_{2b}(z)V_{2b}(0)V_{2b}(1)V_{2b}(\infty)\rangle\nonumber\\+
\langle\Psi_b(z)\Psi_b(0)\Phi_b(1)\Phi_b(\infty)\rangle \langle
V_{2b}(z)V_{2b}(0)V_{2b}(1)V_{2b}(\infty)\rangle\nonumber\\+
\langle\Psi_b(z)\Phi_b(0)\Phi_b(1)\Psi_b(\infty)\rangle \langle
V_{2b}(z)V_{2b}(0)V_{2b}(1)V_{2b}(\infty)\rangle\bigg)\nonumber\\
+\bigg(\langle\chi_b(z)\Phi_b(0)\chi_b(1)\Phi_b(\infty)\rangle
\langle\bar
Y_{2b}(z)V_{2b}(0)\bar Y_{2b}(1)V_{2b}(\infty)\rangle\nonumber\\
+\langle\chi_b(z)\chi_b(0)\Phi_b(1)\Phi_b(\infty)\rangle \langle\bar
Y_{2b}(z)\bar Y_{2b}(0)V_{2b}(1)V_{2b}(\infty) \rangle\nonumber\\
+\langle\chi_b(z)\Phi_b(0)\Phi_b(1)\chi_b(\infty)\rangle \langle\bar
Y_{2b}(z)V_{2b}(0)V_{2b}(1)\bar Y_{2b}(\infty) \rangle\bigg)\label{I42b}\\
+\bigg(\langle\bar\chi_b(z)\Phi_b(0)\bar\chi_b(1)\Phi_b(\infty)\rangle
\langle  Y_{2b}(z)V_{2b}(0) Y_{2b}(1)V_{2b}(\infty) \rangle\nonumber\\
+\langle\bar\chi_b(z)\bar\chi_b(0)\Phi_b(1)\Phi_b(\infty)\rangle
\langle  Y_{2b}(z)Y_{2b}(0) V_{2b}(1)V_{2b}(\infty) \rangle\nonumber\\
+\langle\bar\chi_b(z)\Phi_b(0)\Phi_b(1)\bar\chi_b(\infty)\rangle
\langle  Y_{2b}(z)V_{2b}(0) V_{2b}(1)Y_{2b}(\infty) \rangle\bigg)\nonumber\\
+ \bigg(\langle \Phi_b(z)\Phi_b(0)\Phi_b(1)\Phi_b(\infty)\rangle
\langle W_{2b}(z)V_{2b}(0)W_{2b}(1)V_{2b}(\infty)\rangle\nonumber\\
+ \langle\Phi_b(z)\Phi_b(0)\Phi_b(1)\Phi_b(\infty)\rangle \langle
W_{2b}(z)W_{2b}(0)V_{2b}(1)V_{2b}(\infty)\rangle\,\,\,\nonumber\\ +
\langle \Phi_b(z)\Phi_b(0)\Phi_b(1)\Phi_b(\infty)\rangle \langle
W_{2b}(z)V_{2b}(0)V_{2b}(1)W_{2b}(\infty)\rangle\bigg)\bigg].\nonumber
\end{align}
We now use the conformal block decomposition of the correlation
functions. It is useful to introduce a compact notation. For a
while, we omit some arguments that are easily reconstructed in the
final expressions. In the matter sector,
\begin{align}
&\langle \Phi(z) \Phi(0) \Phi(1) \Phi(\infty) \rangle= c_k |A_k^{(0)}(z)|^2,  \nonumber\\
&\langle \Psi(z) \Phi(0) \Psi(1) \Phi(\infty) \rangle= c_k |A_k^{(1)}(z)|^2,  \nonumber\\
&\langle \Psi(z) \Psi(0) \Phi(1) \Phi(\infty) \rangle= c_k
|A_k^{(2)}(z)|^2, \ \label{AA}\\
&\langle \Psi(z) \Phi(0) \Phi(1) \Psi(\infty) \rangle= c_k
|A_k^{(3)}(z)|^2. \nonumber
\end{align}
Here, the index $k=+,0,-$ corresponds to the three channels in the
degenerate OPE of the field $\Phi_b$ (and also of its super
partners), and we assume summation with respect to $k$. The
coefficients $c_k$ are related to the basic structure constants
(see~\cite{BBSLG}):
\begin{align}
&c_+=C_+^2(b)=\frac{\gamma(1/2+b^2/2)\gamma(-1/2+5b^2/2)}{\gamma(-1/2+3b^2/2)\gamma(1/2+3b^2/2)},\\
&c_0=-\tilde C_0^2(b)=-1,\\
&c_-=C^2_-(b)=-\frac{\gamma(2b^2)\gamma(-1/2+b^2/2)\gamma^2(1/2+b^2/2)}
{b^4\gamma^3(b^2)\gamma(-1+b^2)\gamma(-1/2+3b^2/2)}.
\end{align}
In~\eqref{AA}, $A_k^{(n)}$ denotes the conformal blocks appearing in the
$k$ channel for the given correlation function. Here and hereafter, the
normalization is chosen such that all but the basic combinations $c_k$
are absorbed inside the conformal blocks. In Appendix~A, we recapitulate
some details and explicit constructions concerning conformal blocks.
In the Liouville sector,
\begin{align}
&\langle V(z) V(0) V(1) V(\infty) \rangle= \mathcal{R}\int
\frac{dP}{4\pi} r_{l}(P) |B_l^{(0)}(P,z)|^2, \nonumber\\
&\langle W(z) V(0) W(1) V(\infty) \rangle= \mathcal{R}\int
\frac{dP}{4\pi} r_{l}(P) |B_l^{(1)}(P,z)|^2,  \nonumber\\
&\langle W(z) W(0) V(1) V(\infty) \rangle= \mathcal{R}\int
\frac{dP}{4\pi} r_{l}(P) |B_l^{(2)}(P,z)|^2, \label{BB}\\
&\langle W(z) V(0) V(1) W(\infty) \rangle= \mathcal{R}\int
\frac{dP}{4\pi} r_{l}(P) |B_l^{(3)}(P,z)|^2. \nonumber
\end{align}
Because the correlation functions do not contain the degenerate fields
in the Liouville sector, the index $l$ here assumes summation of the
``even'' and of the ``odd'' conformal blocks in accordance with general
OPE~\eqref{VV}. Again, the normalization leaves only two basic combinations
outside the conformal blocks,
\begin{align}
\mathcal{R} r_0(P)=
\mathbb{C}_{a,a}^{Q/2+iP}\mathbb{C}_{a,a}^{Q/2-iP}, \nonumber\\
\mathcal{R}r_1(P)= \mathbb{\tilde C}_{a,a}^{Q/2+iP}\mathbb{\tilde
C}_{a,a}^{Q/2-iP},
\end{align}
where we separate the factor $\mathcal{R}$ for convenience (it is
independent of $P$) and the parameter $a$ is related to the external
conformal dimension (i.~e. either $b$ or $2b$). All other
correlation functions in~\eqref{I42b} are already not independent.
They are expressed in terms of the same ingredients as in~\eqref{AA}
and \eqref{BB}. For example,
\begin{align}
&\langle \chi(z) \Phi(0) \chi(1) \Phi(\infty) \rangle= c_k A_k^{(1)}(z) A_k^{(0)}(\bar z), \nonumber\\
&\langle \chi(z) \chi(0) \Phi(1) \Phi(\infty) \rangle= c_k A_k^{(2)}(z) A_k^{(0)}(\bar z), \label{AA1}\\
&\langle \chi(z) \Phi(0) \Phi(1) \chi(\infty) \rangle= c_k
A_k^{(3)}(z) A_k^{(0)}(\bar z), \nonumber
\end{align}
and
\begin{align}
\langle \bar Y(z) V(0) \bar Y(1) V(\infty) \rangle= \mathcal{R}\int
\frac{dP}{4\pi} r_{l}(P) B_l^{(0)}(P, z) \bar B_l^{(1)}(P,\bar z), \nonumber\\
\langle \bar Y(z) \bar Y(0) V(1) V(\infty) \rangle=\mathcal{R}\int
\frac{dP}{4\pi} r_{l}(P) B_l^{(0)}(P, z) \bar B_l^{(2)}(P,\bar z), \label{BB1}\\
\langle \bar Y(z) V(0) V(1) \bar Y(\infty) \rangle= \mathcal{R}\int
\frac{dP}{4\pi} r_{l}(P) B_l^{(0)}(P, z) \bar B_l^{(3)}(P, \bar z).
\nonumber
\end{align}
The remaining six correlation functions not written explicitly are
obtained from~\eqref{AA1} and~\eqref{BB1} by complex conjugation.
Using the introduced notation, we can rewrite the integrals under
consideration in the compact forms
\begin{align}
\mathcal{I}_{4}(b)= 2\mathcal{R}\int_{\mathbf{G}} d^2 z \int
\frac{dP}{4\pi} \sum_{l}r_{l}(P) \bigg[
|B_l^{(1)}(P,z)|^2+|B_l^{(2)}(P,z)|^2+|B_l^{(3)}(P,z)|^2
\bigg]\label{I4bcomp}
\end{align}
and
\begin{align}
&\mathcal{I}_{4}(2b)= 2\mathcal{R}\int_{\mathbf{G}} d^2 z \int
\frac{dP}{4\pi} \sum_{k,l}c_k r_{l}(P) \bigg[
|A_k^{(1)}(z)B_l^{(0)}(P,z)+A_k^{(0)}(z)B_l^{(1)}(P,z)|^2\nonumber\\
&+|A_k^{(2)}(z)B_l^{(0)}(P,z)+A_k^{(0)}(z)B_l^{(2)}(P,z)|^2+|A_k^{(3)}(z)B_l^{(0)}(P,z)+A_k^{(0)}(z)B_l^{(3)}(P,z)|^2
\bigg]. \label{I42bcomp}
\end{align}
Bulky expression~\eqref{I42b} has a remarkably compact and clear structure
in terms of the conformal blocks. We again note that in~\eqref{I4bcomp}
and~\eqref{I42bcomp}, we respectively assume the different values of
external conformal dimensions in the Liouville sector $\Delta_b$ and
$\Delta_{2b}$.

\section{The Modular Integral}

It turns out efficient~\cite{basic} to use elliptic transformations
in the integration. We use the standard map
\begin{align}
\tau=i \frac{K(1-z)}{K(z)},
\end{align}
where the complete elliptic integral of the first kind is
\begin{align}
K(z)=\frac{1}{2} \int_0^1 \frac{d t}{y}
\end{align}
and $y^2=t(1-t)(1-z t)$. It can be verified that
\begin{align}
dz=\pi z(1-z)\theta_3^4(q)d\tau,
\end{align}
where
\begin{align}
q=e^{i \pi \tau}
\end{align}
and
\begin{align}
\theta_3(q)=\sum_{n=-\infty}^{\infty}q^{n^2}.
\end{align}
Integral~\eqref{I4bcomp} becomes
\begin{align}
\mathcal{I}_4(b)= 2\pi^2 \mathcal{R}\int
_{-\infty}^{\infty}\frac{dP}{4\pi} \sum_{l}r_{l}(P)
\bigg[\int_{\mathbf{F}}
|z(1-z)\theta_3^4(q)B_l^{(1)}(P,z)|^2d^{2}\tau\nonumber\\+\int_{\mathbf{F}}|z(1-z)\theta_3^4(q)B_l^{(2)}(P,z)|^2d^{2}\tau+
\int_{\mathbf{F}}|z(1-z)\theta_3^4(q)B_l^{(3)}(P,z)|^2d^{2}\tau\bigg],
\label{I4bev2}
\end{align}
where $\mathbf{F}$\textbf{\ }$=\left\{  \left|  \tau\right|
>1;\;\left| \operatorname*{Re}\tau\right|  <1/2\right\}$. Similarly,
for~\eqref{I42bcomp}, we have
\begin{align}\label{I42bev2}
\mathcal{I}_4(2b)=& 2\pi^2 \mathcal{R}\int
_{-\infty}^{\infty}\frac{dP}{4\pi} \sum_{k,l}c_k r_{l}(P)\nonumber\\
\bigg[&\int_{\mathbf{F}}|z(1-z)\theta_3^4(q)(A_k^{(1)}(z)B_l^{(0)}(P,z)+A_k^{(0)}(z)B_l^{(1)}(P,z))|^2d^{2}\tau\\
+&\int_{\mathbf{F}}|z(1-z)\theta_3^4(q)(A_k^{(2)}(z)B_l^{(0)}(P,z)+A_k^{(0)}(z)B_l^{(2)}(P,z))|^2d^{2}\tau\nonumber\\
+&\int_{\mathbf{F}}|z(1-z)\theta_3^4(q)(A_k^{(3)}(z)B_l^{(0)}(P,z)+A_k^{(0)}(z)B_l^{(3)}(P,z))|^2d^{2}\tau
\bigg]. \nonumber
\end{align}
We now define the conformal blocks more explicitly,
\begin{align}
A^{(0)}_{-}(z)=F_{00}^{\text{M}}(0, z),\,\,
\,\,\,A^{(0)}_{0}(z)=F_{01}^{\text{M}}(b, z),\,\, \,\,\,
A^{(0)}_{+}(z)=F_{00}^{\text{M}}(2b, z), \nonumber\\
A^{(1)}_{-}(z)=F_{11}^{\text{M}}(0, z),\,\,
\,\,\,A^{(1)}_{0}(z)=F_{10}^{\text{M}}(b, z),\,\, \,\,\,
A^{(1)}_{+}(z)=F_{11}^{\text{M}}(2b, z), \\
A^{(2)}_{-}(z)=F_{20}^{\text{M}}(0, z),\,\,
\,\,\,A^{(2)}_{0}(z)=F_{21}^{\text{M}}(b, z),\,\, \,\,\,
A^{(2)}_{+}(z)=F_{20}^{\text{M}}(2b, z), \nonumber\\
A^{(3)}_{-}(z)=F_{31}^{\text{M}}(0, z),\,\,
\,\,\,A^{(3)}_{0}(z)=F_{30}^{\text{M}}(b, z),\,\, \,\,\,
A^{(3)}_{+}(z)=F_{31}^{\text{M}}(2b, z).\nonumber
\end{align}

Here, the first argument of the symmetric conformal blocks defines
the internal conformal dimension. The first lower index corresponds
to one of the four basic types of conformal blocks we consider with
respect to the set of external fields (see Appendix~A); the second
index is $0$ if the corresponding block with the given internal
conformal dimension is ``even'' and $1$ if it is ``odd''. In the
Liouville sector,
\begin{align}
B^{(0)}_{0}(z)=F_{00}^{\text{L}}(P, z),\,\,
\,\,\,B^{(0)}_{1}(z)=F_{01}^{\text{L}}(P, z),\nonumber\\
B^{(1)}_{0}(z)=F_{11}^{\text{L}}(P, z),\,\,
\,\,\,B^{(1)}_{1}(z)=F_{10}^{\text{L}}(P, z),\\
B^{(2)}_{0}(z)=F_{20}^{\text{L}}(P, z),\,\,
\,\,\,B^{(2)}_{1}(z)=F_{21}^{\text{L}}(P, z),\nonumber\\
B^{(3)}_{0}(z)=F_{31}^{\text{L}}(P, z),\,\,
\,\,\,B^{(3)}_{1}(z)=F_{30}^{\text{L}}(P, z).\nonumber
\end{align}

For the first integral, the complicated expressions for the
Liouville structure constants give
\begin{align}
\mathcal{R}= \left(\pi  \mu
\gamma(\frac{bQ}{2})b^{1-b^{2}}\right)^{(Q/b-4)}
\Upsilon_b^6(b)\Upsilon_b^2(Q/2)\Upsilon_b^4(Q/2+b).
\end{align}
The special function $\Upsilon_b(x)$ is the standard element of the
LFT (see~\cite{LFT} for the definition and properties). Explicitly,
\begin{align}
\mathcal{R}= &\left(\pi \mu\gamma(\frac{bQ}{2})
b^{1-b^{2}}\right)^{(Q/b-4)}\gamma^4(\frac{bQ}{2})b^{-4b^2}\times\nonumber\\
&\exp\bigg\{\int_0^{\infty}\frac{d t}{t} \bigg[\frac{3(1-
b^2)^2e^{-t}}{2b^2} - \frac{6 \sinh^2((1-b^2)t/(4b))}
{\sinh(t/(2b))\sinh(b t/2)}\bigg] \bigg\},
\end{align}
where we use ``shift'' relations (see~\cite{LFT}) for the last
upsilon function to extract the additional factor
$(\gamma(\frac{bQ}{2}))^{4}$. This allows improving the accuracy of
the comparison with the analytic result, which contains the same
factor with a singularity at $b=1$. The $P$ dependent parts are
\begin{align}
r_{0}(P) &=\frac{P^2\Upsilon_{b}(b\pm iP) \Upsilon_{b}(Q/2\pm
iP)}{\Upsilon_{b}^2(b-Q/4\pm iP/2)\Upsilon_{b}^2(Q/4\pm iP/2)
\Upsilon_{b}^2(b+Q/4\pm iP/2)\Upsilon_{b}^2(3Q/4\pm iP/2)},
\\
r_{1}(P) &=
 \frac{4P^2\Upsilon_{b}(b\pm iP)
\Upsilon_{b}(Q/2\pm iP)}{\Upsilon_{b}^2(3b/2-Q/4\pm
iP/2)\Upsilon_{b}^4(b/2+Q/4\pm iP/2)\Upsilon_{b}^2(1/2/b+Q/4\pm
iP/2)},
\end{align}
where we again use the shift relations to move the arguments of all
upsilon functions inside the strip $[0,Q]$ where the standard
integral representation is applicable and we use the notation
$\Upsilon_{b}(x\pm y)=\Upsilon_{b}(x+y)\Upsilon_{b}(x-y)$
\begin{align}
&r_{0}(P) =P^2\exp\bigg\{\int_0^{\infty}\frac{d t}{t}
\bigg[\frac{(5- 2b^2+5b^4)^2e^{-t}}{2b^2} + \\
&\frac{2 \cos(\frac{P
t}{2})(\cosh(\frac{(1-3b^2)t}{4b})+2\cosh(\frac{(1+b^2)t}{4b})+\cosh(\frac{(3-b^2)t}{4b}))-\cos(P
t)(\cosh(\frac{(1-b^2)t}{2b})+1)-6} {\sinh(t/(2b))\sinh(b
t/2)}\bigg] \bigg\},\nonumber \\
&r_{1}(P) =4P^2\exp\bigg\{\int_0^{\infty}\frac{d t}{t}
\bigg[-\frac{5(1- b^2)^2e^{-t}}{2b^2} + \\
&\frac{2 \cos(\frac{P
t}{2})(3\cosh(\frac{(1-b^2)t}{4b})+\cosh(\frac{3(1-b^2)t}{4b}))-\cos(P
t)(\cosh(\frac{(1-b^2)t}{2b})+1)-6} {\sinh(t/(2b))\sinh(b
t/2)}\bigg] \bigg\}.\nonumber
\end{align}

For the second integral, we analogously find the explicit expressions
\begin{align}
\mathcal{R}= &\left(\pi \mu\gamma(\frac{bQ}{2})
b^{1-b^{2}}\right)^{(Q/b-8)}\gamma^4(b^2)\gamma^4(\frac{bQ}{2})\gamma^4(b^2+\frac{bQ}{2})b^{4(1-6b^2)}\times\nonumber\\
&\exp\bigg\{\int_0^{\infty}\frac{d t}{t} \bigg[\frac{3(1-
b^2)^2e^{-t}}{2b^2} - \frac{6 \sinh^2((1-b^2)t/(4b))}
{\sinh(t/(2b))\sinh(b t/2)}\bigg] \bigg\}
\end{align}
and
\begin{align}
&r_{0}(P) =P^2\exp\bigg\{\int_0^{\infty}\frac{d t}{t}
\bigg[-\frac{(5- 18b^2+37b^4)e^{-t}}{2b^2} + \\
&\frac{2 \cos(\frac{P
t}{2})(\cosh(\frac{(1-7b^2)t}{4b})+2\cosh(\frac{(1+b^2)t}{4b})+\cosh(\frac{(3-5b^2)t}{4b}))-\cos(P
t)(\cosh(\frac{(1-b^2)t}{2b})+1)-6} {\sinh(t/(2b))\sinh(b
t/2)}\bigg] \bigg\},\nonumber \\
&r_{1}(P) =4P^2\exp\bigg\{\int_0^{\infty}\frac{d t}{t}
\bigg[-\frac{(5-26 b^2+37b^4)e^{-t}}{2b^2} + \\
&\frac{2 \cos(\frac{P
t}{2})(\cosh(\frac{(3-7b^2)t}{4b})+2\cosh(\frac{(1-b^2)t}{4b})+\cosh(\frac{(1-5b^2)t}{4b}))-\cos(P
t)(\cosh(\frac{(1-b^2)t}{2b})+1)-6} {\sinh(t/(2b))\sinh(b
t/2)}\bigg] \bigg\}.\nonumber
\end{align}

The conformal blocks can be evaluated effectively using a numerical
algorithm based on the recurrence relations developed
in~\cite{BBNZ,LH1,LH3,VB0,VB1,LH4}. We do not use the elliptic
recursion to construct the necessary conformal blocks here. It turns
out that to attain a convincing accuracy of the results, we need to
know a very few first terms in the $q$-expansion of the conformal
blocks. This information can be obtained directly using the
definition of the conformal blocks in terms of the chain vectors
(see Appendix~A) and the subsequent elliptic transformation.
Nevertheless, the elliptic representation and especially the form of
the prefactor (i.e., the $\Delta$ asymptotic of the conformal
blocks; see~\cite{LH1}) is very useful. It allows verifying the
explicit expressions for the correlation functions by checking the
crossing symmetry requirement, and in particular, fixing up all the
signs, which is difficult to do starting from general principles.

\section{Numerics}

With~\eqref{I4bev2} and~\eqref{I42bev2}, calculating reduces to
numerically integrating several integrals of the general form
\begin{equation}
\int_{\mathbf{F}}|z(1-z)\theta_{3}^{4}(q)\mathcal{F}_P(z)|^2
d^2\tau,\label{formint}
\end{equation}
where $\mathcal{F}_P(z)$ is some Liouville conformal block like
in~\eqref{I4bev2} or some more complicated composite expression like
in~\eqref{I42bev2}. The integrand can be developed as a double power
series in $q$ and $\bar q$ in accordance with the general expansion
\begin{equation}
z(1-z)\theta_{3}^{4}(q)\mathcal{F}_P(z)=\left( 16q\right)^{\alpha}
\sum_{r=0}^{\infty} b_{r}(P)q^{r},\label{Hr}
\end{equation}
where $\alpha$ and the coefficients  $b_{r}$ are defined by the
concrete choice of the function $\mathcal{F}_P(z)$. In each term, we
can integrate in $\tau_{2}=\operatorname*{Im}\tau$ explicitly with
the result in terms of the function
\begin{equation}
\Phi(A,r,l)=\int_{\mathbf{F}}d^{2}\tau\left|  16q\right|
^{2A}q^{r}\bar
q^{l}=\frac{(16)^{2A}}{\pi(2A+r+l)}\int_{-1/2}^{1/2}\cos(\pi(r-l)x)e^{-\pi
\sqrt{1-x^{2}}\left(  2A+r+l\right)  }dx.\label{Phi}%
\end{equation}

For~\eqref{I4bev2}, we have the sum of six integrals of
form~\eqref{formint}, and we obtain the series
\begin{equation}
\mathcal{I}_4(b)=\pi \mathcal{R} \sum_{L=0}^{\infty}\bigg(
A_{L}^{(1,e)}+A_{L}^{(1,o)}+A_{L}^{(2,e)}+A_{L}^{(2,o)}+A_{L}^{(3,e)}+A_{L}^{(3,o)}\bigg),\label{JintL}
\end{equation}
where in the last sum
\begin{align}
&A_{L}^{(1,e)}=\int_{0}^{\infty}r_{2}(P)dP\sum_{k=0}^{L}b_{k}^{(1,e)}(P)b_{L-k}^{(1,e)}(P)\Phi\left(P^{2}/2+Q^{2}/8-1/2,k,L-k\right),\nonumber\\
&A_{L}^{(1,o)}=\int_{0}^{\infty}r_{1}(P)dP\sum_{k=0}^{L}b_{k}^{(1,o)}(P)b_{L-k}^{(1,o)}(P)\Phi\left(P^{2}/2+Q^{2}/8,k,L-k\right),\nonumber\\
&A_{L}^{(2,e)}=\int_{0}^{\infty}r_{2}(P)dP\sum_{k=0}^{L}b_{k}^{(2,e)}(P)b_{L-k}^{(2,e)}(P)\Phi\left(P^{2}/2+Q^{2}/8-1/2,k,L-k\right),\nonumber\\\label{Alevel}
&A_{L}^{(2,o)}=\int_{0}^{\infty}r_{1}(P)dP\sum_{k=0}^{L}b_{k}^{(2,o)}(P)b_{L-k}^{(2,o)}(P)\Phi\left(P^{2}/2+Q^{2}/8,k,L-k\right),\\
&A_{L}^{(3,e)}=\int_{0}^{\infty}r_{1}(P)dP\sum_{k=0}^{L}b_{k}^{(3,e)}(P)b_{L-k}^{(3,e)}(P)\Phi\left(P^{2}/2+Q^{2}/8-1,k,L-k\right),\nonumber\\
&A_{L}^{(3,o)}=\int_{0}^{\infty}r_{2}(P)dP\sum_{k=0}^{L}b_{k}^{(3,o)}(P)b_{L-k}^{(3,o)}(P)\Phi\left(P^{2}/2+Q^{2}/8-1/2,k,L-k\right).\nonumber
\end{align}
Each term in~\eqref{Alevel} is suppressed by a factor
$\max_{\mathbf{F}}\left| q\right|  ^{2L}$, and the series in $L$
converges very rapidly in practice. We found that it suffices to sum
up to $L=4$ to reach the three- to four-digit precision (see
Table~1). In the case $(1,3)$, we numerically integrate in the same
way although we now have 18 integrals of form~\eqref{formint}. In
this case summing up to $L=4$ we were able to reach the two- to
three-digit precision (see Table~2). In Figs.~1 and~2, the results
of numerically evaluating integrals~\eqref{I1analit}
and~\eqref{I2analit} are shown as circles while the lines correspond
to the exact result.
\medskip

\textbf{Acknowledgments} The author thanks the LPTA of University
Montpellier II, for the warm hospitality and stimulating scientific
air. Special gratitude is extended to V.~Fateev and A.~Neveu for
encouraging the interest in this work. He is grateful to A.~Belavin
for the useful discussions. This work was supported by the Russian
Foundation for Basic Research (Grant No. 08-01-00720) and also by
the RBRF-CNRS project PICS-09-02-91064. Part of the calculations
were performed while visiting the Mathematical Department of Kyoto
University in January 2008. He acknowledges the hospitality of this
division and personally T.~Miwa.

\begin{center}%
\begin{table}[tbp] \centering
\begin{tabular}
[c]{|c|c|c|}\hline $b$ & $\Sigma^{(1,1)}(b)$ num. & $\Sigma^{(1,1)}(b)$ exact \\
\hline
0.999 & 0.9959 & 0.9960 \\
0.95 &  0.8049 & 0.8050 \\
0.85 &  0.4450 & 0.4450 \\
0.80 &  0.2799 & 0.2800 \\
$1/\sqrt{2}$ & 0.0001 & 0 \\
0.65 & 0.1555 & 0.1550 \\
0.60 & 0.2877 & 0.2800\\ \hline
\end{tabular}
\caption{Numerical data for $\Sigma^{(1,1)}(b)$ at $\mu=1$.}
\end{table}%
\begin{table}[tbp] \centering
\begin{tabular}
[c]{|c|c|c|}\hline $b$ & $\Sigma^{(1,3)}(b)$ num. &
$\Sigma^{(1,3)}(b)$ exact \\ \hline
0.71 & 0.0309 & 0.0246\\
0.69 & -0.1377 & -0.1434\\
0.67 & -0.3030 & -0.3066\\
0.65 & -0.4623 & -0.4650\\
0.63 & -0.6159 &-0.6186\\
0.61 &-0.7644 &-0.7674\\
0.59 &-0.9096 &-0.9114\\
0.57& -0.9699 & -0.9747\\
0.55 & -0.9060 &-0.9075\\
0.53 & -0.8409 &-0.8427 \\
0.51 & -0.7791 &-0.7803\\
0.49 & -0.7197 & -0.7203\\
0.47 & -0.6621 &  -0.6627 \\
0.45 & -0.6069 & -0.6075\\
0.43& -0.3342 & -0.3282 \\
0.41&-0.0384 & -0.0258\\
0.39& 0.1827 & 0.2622\\ \hline
\end{tabular}
\caption{Numerical data for $\Sigma^{(1,3)}(b)$ at $\mu=1$.}
\end{table}%
\end{center}
\begin{figure}[!htb]
\centerline{\includegraphics[angle=0,clip=true]{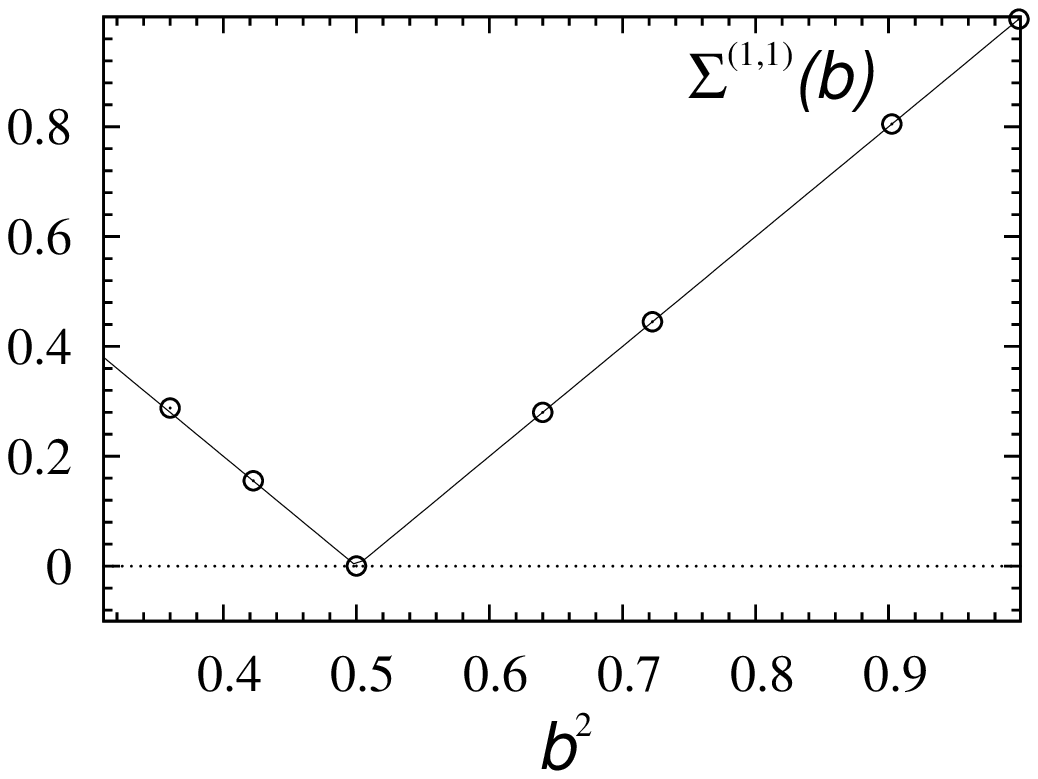}}
\caption{Direct numerical evaluation of reduced
integral~\eqref{I1analit} (circles) versus the exact formula
(continuous line). } \label{fig:I4b}
\centerline{\includegraphics[angle=0,clip=true]{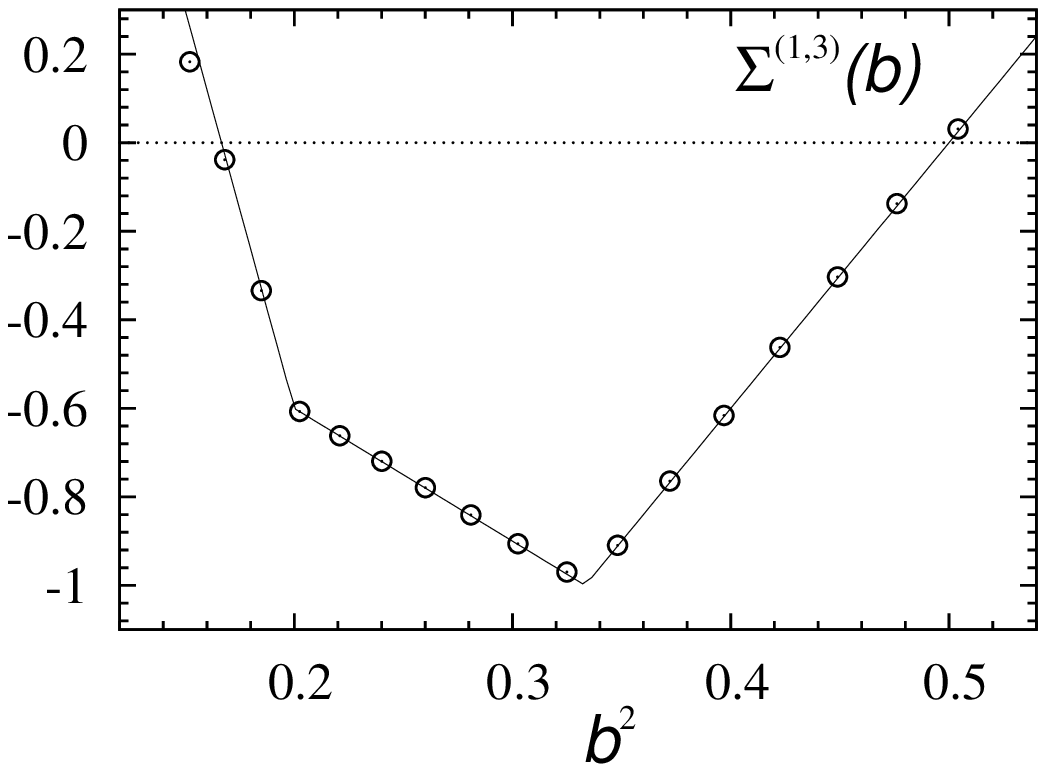}}
\caption{Direct numerical evaluation of reduced
integral~\eqref{I2analit} (circles) versus the exact formula
(continuous line).} \label{fig:boostrap}
\end{figure}


\appendix{\bf Appendix A. Conformal Blocks and the Chain Vectors}\\
For definiteness, we use the notation for the Liouville sector,
though the results concerning the conformal block are universal
(i.e., are independent of the sector). Schematically, the
contribution of the given conformal family in the four basic OPE can
be written as
\begin{align}
V_1(z) V_2(0) &= z^{\Delta-\Delta_1-\Delta_2} \sum_N z^N
|N\rangle_{12},\nonumber\\
W_1(z) V_2(0) &= z^{\Delta-\Delta_1-\Delta_2-1/2} \sum_N z^N
\widetilde{|N\rangle}_{12},\nonumber\\
V_1(z) W_2(0) &= z^{\Delta-\Delta_1-\Delta_2-1/2} \sum_N z^N
\widetilde{\widetilde{|N\rangle}}_{12},\\
W_1(z) W_2(0) &= z^{\Delta-\Delta_1-\Delta_2-1} \sum_N z^N
\widetilde{\widetilde{\widetilde{|N\rangle}}}_{12},\nonumber
\end{align}
where the so-called chain vectors
$|N\rangle,\widetilde{|N\rangle},\widetilde{\widetilde{|N\rangle}},\widetilde{\widetilde{\widetilde{|N\rangle}}}$
(with positive integer or half-integer $N$) are the $N$th-level
descendent contribution of the intermediate state with the conformal
dimension $\Delta$ appearing in the given operator product
expansion. The chain vectors are completely determined by the
superconformal symmetry. Namely, the superconformal constraints lead
to the recurrence relations
\begin{equation}
\begin{cases}
G_k|N\rangle_{12}={\widetilde{|N-k \rangle}}_{12},\\
G_k{\widetilde{|N\rangle}}_{12}=
[\Delta+2k\Delta_1-\Delta_2+N-k]|N-k\rangle_{12}\end{cases}
\label{chain1}
\end{equation}
for $k>0$. And
\begin{align}
\begin{cases}
G_k \widetilde{\widetilde{|N\rangle}}_{12}=
\widetilde{\widetilde{\widetilde{|N-k\rangle}}}_{12}+2\Delta_2 \delta_{k,1/2}|N-k\rangle_{12},\\
G_k
\widetilde{\widetilde{\widetilde{|N\rangle}}}_{12}=[\Delta+2k\Delta_1-(\Delta_2+1/2)+N-k]
\widetilde{\widetilde{|N-k\rangle}}_{12}-2\Delta_2
\delta_{k,1/2}\widetilde{|N-k\rangle}_{12}\label{chain2}
\end{cases}
\end {align}
for $k>0$. The normalization of the chain vectors chosen in this
text is determined by the requirements
\begin{align}
|0\rangle=1,\,\,\,\,\, \widetilde{|0\rangle}=1,\,\,\,\,\,
\widetilde{\widetilde{|0\rangle}}=-1,\,\,\,\,\,
\widetilde{\widetilde{\widetilde{|0\rangle}}}=(\Delta-\Delta_1-\Delta_2).
\end {align}

Relations~\eqref{chain1} and~\eqref{chain2} are equivalent to the
linear problem for the coefficients determining the chain vectors in
terms of the Virasoro basis vectors of the same level. These systems
can be solved numerically up to a rather high level.

The necessary $s$-channel superconformal blocks are defined via the
expansions
\begin{align}
&\mathcal{F}_{\text{e,o}}\left(\left.\begin{array}[c]{cc}%
a_1&a_3\\a_2&a_4\end{array}\right|\Delta\,\bigg|\,z\right)=
z^{\Delta-\Delta_1-\Delta_2}\sum_{N\in
\mathbb{Z},\mathbb{Z}/2}z^N{}_{12} \langle N|N\rangle_{34},
\label{ConfBlockDef0}
\\
&\mathcal{F}_{\text{e,o}}\left(\left.\begin{array}[c]{cc}%
\hat a_1&\hat
a_3\\a_2&a_4\end{array}\right|\Delta\,\bigg|\,z\right)=
z^{\Delta-\Delta_1-\Delta_2-1/2}\sum_{N\in
\mathbb{Z},\mathbb{Z}/2}z^N{}_{12} \wtbrN|\wtktN_{34},
\label{ConfBlockDef1}
\\
&\mathcal{F}_{\text{e,o}}\left(\left.\begin{array}[c]{cc}%
\hat a_1&a_3\\\hat
a_2&a_4\end{array}\right|\Delta\,\bigg|\,z\right)=
z^{\Delta-\Delta_1-\Delta_2-1}\sum_{N\in
\mathbb{Z},\mathbb{Z}/2}z^N{}_{12}\wtbrNNN|N\rangle_{34},
\label{ConfBlockDef2}
\\
&\mathcal{F}_{\text{e,o}}\left(\left.\begin{array}[c]{cc}%
\hat a_1& a_3\\a_2&\hat
a_4\end{array}\right|\Delta\,\bigg|\,z\right)=
z^{\Delta-\Delta_1-\Delta_2-1/2}\sum_{N\in
\mathbb{Z},\mathbb{Z}/2}z^N{}_{12} \wtbrN|\wtktNN_{34}.
\label{ConfBlockDef3}
\end{align}
In the main text, we use the brief notation
\begin{align}
F_{00}(\Delta, z)=\mathcal{F}_{\text{e}}\left(\left.\begin{array}[c]{cc}%
a&a\\a&a\end{array}\right|\Delta\,\bigg|\,z\right),\,\,\,\,\,\,
F_{01}(\Delta, z)=\mathcal{F}_{\text{0}}\left(\left.\begin{array}[c]{cc}%
a&a\\a&a\end{array}\right|\Delta\,\bigg|\,z\right),
\end{align}
\begin{align}
F_{10}(\Delta, z)=\mathcal{F}_{\text{e}}\left(\left.\begin{array}[c]{cc}%
\hat a&\hat
a\\a&a\end{array}\right|\Delta\,\bigg|\,z\right),\,\,\,\,\,\,
F_{11}(\Delta, z)=\mathcal{F}_{\text{0}}\left(\left.\begin{array}[c]{cc}%
\hat a&\hat a\\a&a\end{array}\right|\Delta\,\bigg|\,z\right),
\end{align}
\begin{align}
F_{20}(\Delta, z)=\mathcal{F}_{\text{e}}\left(\left.\begin{array}[c]{cc}%
\hat a& a\\\hat
a&a\end{array}\right|\Delta\,\bigg|\,z\right),\,\,\,\,\,\,
F_{21}(\Delta, z)=\mathcal{F}_{\text{0}}\left(\left.\begin{array}[c]{cc}%
\hat a&a\\\hat a&a\end{array}\right|\Delta\,\bigg|\,z\right),
\end{align}
\begin{align}
F_{30}(\Delta, z)=\mathcal{F}_{\text{e}}\left(\left.\begin{array}[c]{cc}%
\hat a& a\\
a&\hat a\end{array}\right|\Delta\,\bigg|\,z\right),\,\,\,\,\,\,
F_{31}(\Delta, z)=\mathcal{F}_{\text{0}}\left(\left.\begin{array}[c]{cc}%
\hat a& a\\ a&\hat a\end{array}\right|\Delta\,\bigg|\,z\right).
\end{align}

\end{document}